\documentclass[12pt,preprint]{aastex}

\usepackage{lscape}
\usepackage{epstopdf}

\shorttitle{Spectroscopy of Short Period Recurrent Nova}

\shortauthors{Sion, Wilson, Godon, Starrfield, Williams, Darnley}

\begin{document}

\title{HST FUV SPECTROSCOPY OF THE SHORT ORBITAL PERIOD RECURRENT NOVA CI AQL: IMPLICATIONS FOR WHITE DWARF MASS EVOLUTION}

\author{Edward M. Sion}
\affil{Department of Astrophysics and Planetary Science, Villanova University, Villanova, PA 19085, USA} \email{edward.sion@villanova.edu}
 
\author{R. E. Wilson}
\affil{Astronomy Department, University of Florida, Gainesville, FL 32611, USA}  \email{rewilson@ufl.edu}

\author{Patrick Godon\altaffilmark{1}}
\affil{Department of Astrophysics and Planetary Science, Villanova University, Villanova, PA 19085, USA} \email{Patrick.Godon@villanova.edu}

\author{Sumner Starrfield}
\affil{School of Earth and Space Exploration, Arizona State University, Tempe, AZ 85287, USA} \email{sumner.starrfield@gmail.com}

\author{Robert E. Williams}
\affil{Space Telescope Science Institute, 3700 San Martin Drive, Baltimore, MD 21218, USA} \email{wms@stsci.edu}

\author{M.J.Darnley}
\affil{Astrophysics Research Institute, Liverpool John Moores University, IC2 Liverpool Science Park, Liverpool, L3 5RF, UK}\email{M.J.Darnley@ljmu.ac.uk}

\altaffiltext{1}{Henry A. Rowland Department of Physics and Astronomy, The Johns Hopkins University, Baltimore, MD 21218, USA}

\clearpage

\begin{abstract}

An HST COS Far UV spectrum (1170 \AA\ to 1800 \AA ) was obtained for
the short orbital period recurrent novae (T Pyxidis 
 subclass), CI Aquilae. CI Aql is the only classical CV known to have two eclipses of sensible depth per orbit cycle and also have pre- and post-outburst light curves that are
 steady enough to allow estimates of mass and orbital period changes. Our FUV
 spectral analysis with 
 model accretion disks and NLTE high gravity photospheres, together with the
 Gaia parallax, reveal CI Aql's FUV light is dominated by an optically
thick accretion disk with an accretion rate of the order of 
$4\times 10^{-8}$ $M_\sun/yr$. 
Its database of light curves, radial velocity curves, and eclipse timings is among the best for any CV. Its orbit period ($P$),
 $dP/dt$, and reference time are re-derived via simultaneous analysis of
 the three data types, giving a dimensionless post-outburst $dP/dt$ of
 $-2.49\pm 0.95\times 10^{-10}$. Lack of information on loss
 of orbital to rotational angular momentum leads to some uncertainty in the translation of
 $dP/dt$ to white dwarf mass change rate, $dM_1/dt$, but within the modest
 range of $+4.8\times 10^{-8}$
 to $+7.8\times 10^{-8}$ $M_\sun /yr$. The estimated white dwarf mass change through outburst for CI Aql,
 based on simple differencing of its pre- and post outburst orbit period,
 is unchanged from the previously published $+5.3 \times 10^{-6} M_\sun$.
 At the WD's estimated mass increase rate, it will
 terminate as a Type Ia supernova within 10 million years.

\end{abstract}

\section{Introduction: the Short Orbital Period Recurrent Nova CI Aql}

Cataclysmic variables (CVs) are close binaries comprised of a main sequence, sub-giant, 
or giant star that fills its Roche-lobe and transfers gas to a white dwarf (WD) via an accretion disk if the WD is only weakly magnetic or via a magnetically channeled accretion column if the WD is strongly magnetic. When a critical mass of hydrogen-rich gas accumulates on the white dwarf, an explosive thermonuclear runaway(TNR) is triggered, identified as a classical nova. 
The recurrent novae (RNe) are a subset 
of CVs that have suffered more than one recorded (TNR) outburst on recurrence 
timescales of a year to a century. Recent reviews on the properties and basic 
parameters of RNe can be found in \citet{anupama,anupama13} and \citet{schaefer}. 
Their short recurrence times require both a massive WD  
and a high accretion rate \citep{starrfield,yaron}.
These two conditions are required for an accreting WD to explode as a Type Ia supernova (SN Ia),
which is why RNe are among the best possibilities for a solution 
to the longstanding SN Ia progenitor problem \citep{pagnotta}.) This single degenerate scenario and the double-degenerate merger scenario 
\citep{iben84,webbink} both remain viable pathways to SN Ia explosions.

A small subclass of 
RNe (T Pyx, IM Nor, CI Aql) have short orbital periods (relative to other recurrent novae) and main sequence (MS) donor companions. All other RNe have subgiant or red giant donors and long orbital periods (days to years). The hallmark of these three objects is their slow optical decline timescales compared to other recurrent novae. However, estimates of their ejecta
masses, ejection velocities and soft-X-ray runoff times are comparable to other recurrent novae and fast classical novae. This led \citet{caleo} to propose that their slower optical declines can be explained as gas which is transferred into the Roche lobe of the WD by the donor star in the first few days of the nova explosion, blocking the radiation from ionizing the ejecta and increasing the optical decline timescale. Whether or not this explains their distinctive slow declines requires further exploration.

The accretion disks and accreting white dwarfs in CVs (including RNe) have the peaks of their spectral energy distributions in the FUV and are best studied there. FUV observations 
can yield accurate accretion rates, and if the WD is exposed, also its surface temperature and possibly its rotation rate and chemical abundances. Due to its   faintness, CI Aql 
had not previously been observed in the far ultraviolet (FUV). Therefore, we requested 
Hubble Space Telescope (HST) observations of CI Aql with the Cosmic Object 
Spectrograph (COS) and carried out an analysis of the HST COS FUV spectrum using the parameters in 
Table \ref{params}, including the newly available Gaia distances.

\citet{iijima} showed that CI Aql's optical spectral evolution during the 
2000 outburst resembled those of T Pyx type RNe's. \citet{lederle} (hereafter LK) and \citet{mennickent} 
(hereafter MH) found that
CI Aql shows eclipses on an orbital period of 14.8 $hr$, with an evolved (MH) or main sequence (LK) donor star.
Its optical quiescence spectrum is 
very different from those of typical quiescent novae. The spectrum shows weak emission 
lines due to HeII and, in addition, the C III-N III complex on a reddened continuum \citep{anupama13}. 

CI Aql was 
detected as a soft X-ray source 14 and 16 months after the 2000 
outburst \citep{greiner}. From radial velocities (hereafter RV), \citet{sahman} found the 
mass of the WD to be $1.00 \pm 0.14 M_\sun$ and the mass of 
the donor star to be $2.32 \pm 0.19 M_\sun$ (see Table \ref{params}).
They estimate the secondary's radius to be $2.07 \pm 0.06$ $R_\sun$, 
implying that it is a slightly evolved early A-type star. The high mass 
ratio of $M_2/M_1=q = 2.35 \pm 0.24$ and the high secondary-star mass implies 
that the mass transfer occurs on a thermal time-scale. \citet{sahhill}
suggest that CI Aql may be evolving into a persistent supersoft x-ray source, 
and may eventually explode as an SN Ia. Models also suggest it may 
ultimately explode as a SN Ia, possibly within 10 Myr \citep{sahman}.
Moreover, \citet{wilshoney} (hereafter WH) found that both the 2000 post-outburst rate of period change, $dP/dt$,
and $\Delta M_1$ (pre- vs. post-outburst $M_1$ change) indicate that the WD is growing in mass. 

Whether the white dwarf in a recurrent nova increases or decreases its mass with each nova 
outburst is key to understanding whether they are progenitors of Type Ia supernovae. Given the critical importance of this question,
we extend WH's analysis of light curves and eclipse timings to include
also RV curves and report the results in later sections.

CI Aql's post-outburst ephemeris ($HJD_0$, $P$, $dP/dt$) is derived in \S\ref{ephemderive} by simultaneous solution of 
all light curves, RV curves, and eclipse timings that could be found (See \S\ref{ephem} for an overview). 
The algorithm \citep{wilson14} iteratively revises the observational weights to 
ensure properly balanced influence of the three data types. Eclipse timings can cover otherwise
blank epochs, but most CI Aql ephemeris information comes from the light and RV
curves. The simultaneous solutions improve the accuracy of $dP/dt$. Then application of eqn. 5
of WH, derived in their \S 5.1, yields $dM_1/dt$ with improved accuracy,
after adoption of a plausible range for conversion of orbital to rotational angular momentum.
Accretion rates ($dM_1/dt$) are derived here from the CI Aql FUV spectrum for the first time and compared
with those for T Pyx \citep{godon18}.
Disk modeling codes are described and the results summarized in \S\ref{theodisk} and \S\ref{fuv}, 
respectively, along with implications for this subclass of RNe.
We also looked for spectroscopic evidence of wind outflow during quiescence 
in the form of P Cygni line profiles and blue-shifted absorption features. 
Spectroscopic detection of the underlying white dwarfs is unlikely, due to the anticipated bright accretion disk.
Details of CI Aql HST COS observations are in \S\ref{hstcos}.

\section{$dP/dt$ and $dM_1/dt$ -- Making the Most of Timing Data} \label{ephem}

A central issue regarding the decades-old and currently undecided conjecture that RNe
are precursors to SN Ia events is whether the WD typically gains or loses
mass over a full nova outburst. Crucial to this point
is whether the orbital period, $P$, increases or decreases because of the nova event, although that may not
be easy or even possible to decide where the pre-explosion observations are sketchy
or absent. Since the WD mass surely decreases in the initial part of the explosion, an
overall increase requires substantially positive $dM_1/dt$ at a subsequent time before the outburst. 
One possible mechanism is 
additional mass accretion during the common envelope stage of the outburst when the 
white dwarf remains bloated following the dynamical ejection \citep{sion14}.
Eqn. 5 of WH quantifies how to compute $dM_1/dt$ from
$dP/dt$ during intervals when there is no mass loss from the system (conservative
case). 

 Traditionally, almost all binary system $dP/dt$ estimates are from eclipse timings, but we 
apply a relatively new and more accurate technique that needs no timing 
estimates \citep{wilson05,wilson06,wilson14}. The traditional way operates 
in two steps -- first measure eclipse times, then fit an ephemeris to the measures. 
Each step has its own errors. The new way operates directly with original 
photometry and RVs, allows step 1 to be
bypassed, and completely eliminates errors in eclipse time estimates.\footnote{Eclipse timings 
often do exist, usually without the original photometry, and can be added to the input stream \citep{wilson14}, 
as is the case with CI Aql, further improving
derived reference time, period, and $dP/dt$.}
Although the ephemeris algorithm streamlines the
process and does save work, the important advance is in accuracy, by elimination of an 
error source. The result is a more accurate ephemeris, including a $dP/dt$ term that leads to
$dM_1/dt$.
An important development is that \citet{sahman} now have RVs of both CI Aql 
components\footnote{The \cite{sahman} star 1 RVs are actually from the wings of emission lines
that presumably originate in the inner disk, close to the white dwarf, and are assumed to
track the spectroscopically invisible white dwarf's orbital motion.} that have put
knowledge of the component star masses on a firm footing. The newly derived masses are very different
from those made in the absence of velocity information.  

\subsection{CI Aql Ephemeris and Masses from Light Curves, Radial Velocities, and Eclipse Timings} \label{ephemderive}

Analysis of CI Aql RVs has been carried out by \cite{sahman} in terms of a traditional point source
model, with an intricate multi-step Monte Carlo solution process. The reader is referred to 
\S 3.9 of \cite{sahman} for their step by step strategy and specifics. 
Our solution is by a differential corrections algorithm ($DC$) that
has been optimized in various ways since its initial publication \citep{wils71}, 
for example, by introduction of simultaneous multi-datatype solutions,
three kinds of data weighting, direct distance estimation, and several ways to improve convergence. 
Light curves, RVs, and eclipse timings exist for CI Aql\footnote{The CI Aql RVs have been 
received from D. Sahman. The timings are from \citet{schaefer2}.}
so combinations of those datatypes can be in the 
input data stream for simultaneous post-outburst $dP/dt$ computation.
With only RVs and eclipse timings as input, 179 Sahman, et al. 
RVs\footnote{Three donor star outlier points among the \cite{sahman} RVs were removed from the
input to our RV analysis. They are HJD 2452802.5963722, RV2 -227.54; HJD 2452802.6593722, RV2 -82.537;
and HJD 2452802.666421, RV2 +209.647.} and 45 Schaefer timings were entered.
Insertion of the 1493 light curve points is not a viable option with a point source model 
since no computed eclipses or proximity effects will match those observed -- the computed 
light curves will be flat. If other than ephemeris parameters are to
be measured, then a similar argument can be made against any model that lacks a disk, since
the light curves, although not flat, will not correspond to a disk model.

Parameters directly at issue are the mass ratio and orbit size (i.e. $a$, the 
orbital semi-major axis), since these are determined well from double-lined spectra and jointly produce
the individual star masses via Kepler's Third Law.
Those quantities are likely to be corrupted by the light curves in simultaneous light/RV no-disk 
solutions, as the diskless geometry is wrong for a CV.
However the published light curves do have ephemeris information, so the issue becomes -- can one tap
into that information by inclusion of the light curves \textit{without damage} to the mass ratio and orbit size results? 
A modest step in the interest of maximizing ephemeris information via insertion of light curves can be
a two part process with the stars made very small (essentially a point source model) 
and the RV curves and timings solved for parameters [$a$,$M_2/M_1$].
The light curves are then included with only [$HJD_0$, $P_0$, $dP/dt$] as output, keeping [$a$,$M_2/M_1$] fixed.
Solutions of this kind with $dP/dt$ adjusted and also fixed at zero were carried out in a few minutes
of machine time.

\subsection{Slightly Changed Ephemeris and Masses}

As we now have the Sahman, et al. RVs, the solution for $P$, $dP/dt$, reference time, and a few 
other parameters from WH was done again with the RVs as additional input. 
Results are in Table \ref{rezults}, with non-zero post-eruption $dP/dt$ now a $2.6\sigma$ 
result compared to the previous $2.4\sigma$ and again with negative sign. 
The derived post-eruption period is the same as in WH, within its uncertainty.
With light curves suppressed, $dP/dt$ was zero within its uncertainty ($+0.9\pm3.1 \times 10^{-10}$),
which is not surprising since all RVs were taken within 2 days and lie within the 
time base of the eclipse timings that produced no measurable $dP/dt$ when analyzed separately. 
This outcome shows that light curves are not just helpful but necessary at this time 
for meaningful estimation of $dP/dt$ in CI Aql.
$dP/dt$ was assumed to be zero for the solution with light curves omitted.\footnote{The
systemic $V_\gamma$ of 42 km $sec.^{-1}$ in Sahman, et al.s Table 3
appears to be a misprint since the horizontal lines that mark $V_\gamma$ in their Fig's 6 and 9
are at $V_\gamma\approx{+4}$ $km$ $sec^{-1}$.}

\subsection{The Rate of Post-outburst White Dwarf Mass Change}

A mis-typed sign has been noticed in the earlier numerical evaluation by WH of 
eqn. 5 for $dM_1/dt$ (the programmed sign was plus [+] instead of the correct minus [-]), in the term involving $dJ/dM_1$.
Their Figure 5 thereby needs revision and an erratum will be published. The figure's purpose
was to quantify the effect on ephemeris-based $dM_1/dt$ of orbital angular momentum ($J_{orb}$) loss from conversion
to rotational angular momentum. Most of the computed $dM_1/dt$'s are now somewhat smaller, following this fix,
although all remain substantially positive for parameters similar to those of CI Aql and any negative $dP/dt$.  
Estimated $dM_1/dt$ now ranges from $+7.8\times 10^{-8}$ $M_\sun /yr$ for negligible loss of orbital angular momentum 
to rotation, down to $+4.8\times 10^{-8}$ $M_\sun /yr$ for maximum plausible loss of $J_{orb}$.
The overall result remains that conversion of orbital to rotational angular momentum is not sufficient
to call the order of estimated (post-eruption) white dwarf mass accretion rate into question. 

\subsection{White Dwarf Mass Change Estimation by Simple Period Differencing} \label{simple}

WH stress that estimated period change (post-outburst minus 
pre-outburst, $\Delta P$) accuracy is limited by uncertainty in pre-outburst $P$,
due to a relatively unpopulated pre-outburst database. That is still true
since we have no new pre-outburst observations. Addition of RVs to the input stream did not significantly change
the post outburst period. With no further 
post-outburst data except for the \cite{sahman} RVs, $\Delta P$'s estimated value remains
as in WH, again being $-2.0\pm 1.4 \times 10^{-6}$ days. $\Delta M_1$ also
is unchanged at $+5.3\pm 3.7 \times 10^{-6} M_\sun$, as it scales with $\Delta P$ while the other 
relevant parameters are nearly the same as before. Although this $\Delta M_1$ result differs from zero by 
only $1.4 \sigma$, the pre- and post-outburst periods upon which it is based are respectively from
well before and well after the outburst of early 2000, so they have as much independence from outburst-induced light
curve disturbances as existing data can provide. 
A set of lightcurves over an interval of about one year in 2001/2002 is illustrated in LK's Fig. 8.
They estimated from overall system brightness that CI Aql's return to quiescence following 
the 2000 outburst occurred in February/March, 2002.
Since the 2001/2002 observations
have a short time base of only 13 months, lie mostly or entirely within the disturbed interval following outburst,
and show much larger asymmetries than those of pre- or post-outburst times,
we shall not attempt to fit them into the overall picture.
LK also utilized pre-outburst photometric coverage from 1991 to 1996 (conveyed by R.K Honeycutt prior to publication), 
from which they estimated CI Aql's pre-outburst $P_{orb}$ to be $0.6183609\pm0.0000009$ by 
the phase dispersion minimization (PDM) algorthm \citep{stellingwerf}. This period 
may seem to differ considerably from that measured by WH \textit{from the same data} 
but the difference is only about half of the rather large statistical uncertainty.

\section{Spectral Analysis} 

\subsection{HST COS Observation of CI Aql} \label{hstcos}

The HST COS spectrum of CI Aql was obtained on November 2, 2017 at 6:36 UT with an exposure 
time 6198.816 s in TIME-TAG mode with the G140L disperser centered at 1105~\AA\ 
(segment B is turned off) through the PSA aperture configured to COS/FUV.  
The COS G140L grating covers 
a wider wavelength range than STIS G140L so our spectra have about 2.5 times better
resolution than STIS. The COS spectrum extends from 1121~\AA\  
to 2148~\AA. However, due to low S/N at the edges of the segment, we ignore the regions 
shorter than 1150~\AA\ and longer than 1900~\AA.

\begin{figure}[h!]
\vspace{-3.5cm}
\includegraphics[width=1.0\textwidth, angle =-00 ]{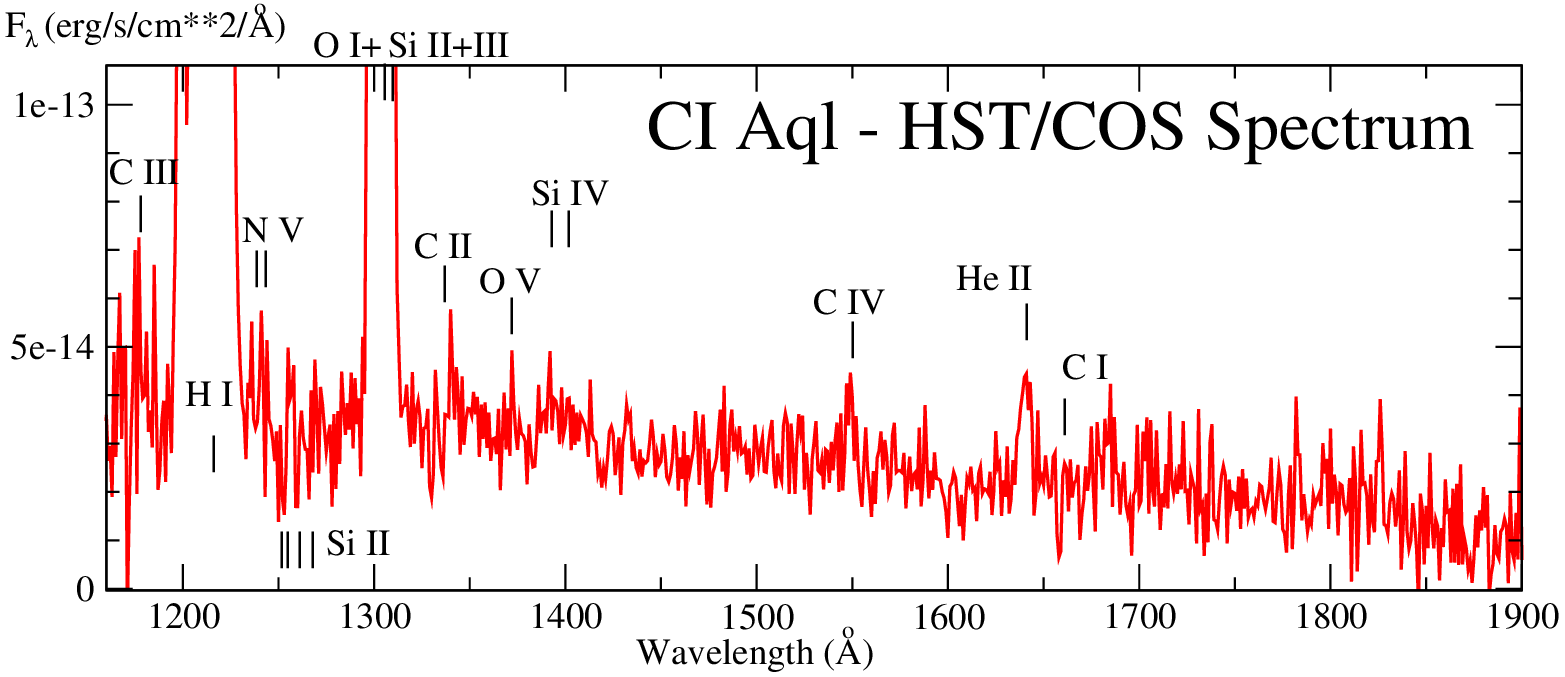}
\vspace{-4.5cm}
\caption{The HST COS spectrum of CI Aql with the strongest line features identified. Strong emission is seen
due to $Ly_\alpha$ and strong O\,{\sc i} + Si\,{\sc iii} (1300A). Relatively weak 
emission features appear due to C\,{\sc iii} (1175), N\,{\sc v} (1238, 1242), C\,{\sc ii},
(1335), Si\,{\sc iv} (1394, 1402), O\,{\sc v}(1371), C\,{\sc iv} (1548, 1550), He\,{\sc ii} (1640), 
and Al II (1670).}
\label{ciaql_id}
\end{figure}

Fig. \ref{ciaql_id} shows the CI Aql spectrum, with a strong 
continuum that rises toward shorter wavelengths. The strongest emission lines are 
$Ly_\alpha$ and O\,{\sc i} + Si\,{\sc iii} (1300~\AA).
Weaker emission lines are seen due to C\,{\sc iii} (1175~\AA), 
C\,{\sc iv} (1550~\AA), He\,{\sc ii} (1640~\AA), 
Al\,{\sc ii} (1670~\AA), and weak O\,{\sc v} (1371~\AA).
 
\subsection{Theoretical Disk and Photosphere Models} \label{theodisk}
 
We use the suite of codes
TLUSTY, SYNSPEC and DISKSYN \citep{hubeny88,hubeny95}, 
for constructing grids of accretion disk models and NLTE high
gravity photosphere models. 
Our disk models, including non-standard disks, are described in \citet{godon17}.
We also adopted model accretion disks from the grid of solar composition 
optically thick, steady state disk 
models (``the standard disk model'') 
of \citet{wade}. In these accretion disk models, the 
outermost disk radius, $R_{\rm out}$, is chosen so that 
$T_{\rm eff}$ ($R_{\rm out}$) is near 10,000K since disk annuli beyond this 
point, which are cooler zones with larger radii, would provide only a very 
small contribution to the mid and far UV disk flux. For the disk models, unless otherwise specified,
we selected every combination of $\dot{M}$, inclination, and white dwarf 
mass to fit the data:  
the inclination angle i = 18, 41, 60, 75 and 81 
degrees, $M_{\rm wd}$ = 0.80, 1.03, 1.21 $M_{\odot}$ and 
$\log (\dot{M})$ ($M_{\odot}$/yr) = -8.5, -9.0, -9.5, -10.0, -10.5. 
For the WD models, we constructed solar composition
WD stellar photospheres with temperatures 
from 12,000 K to 60,000 K in steps of 1,000K to 5,000 K, and with 
effective surface gravity corresponding to the WD mass of the accretion disk model. 
We adopt a projected rotation rate $V_{\rm rot}\sin(i) $ of 200 km/s.
We carried out synthetic spectral fitting with a combination of disks 
and photospheres to model the HST spectra.
 
The fitting errors are mainly due to uncertainties in the WD mass, $M_{\rm wd}$,
and distance, $d$. Similar errors (5-10\% in $T_{eff}$ , $\approx$ 0.5 in log(g))
are obtained if either $M_{\rm wd}$ or $d$ are completely unknown. For accretion disk
spectral fits, if either $d$ or $M_{\rm wd}$ is unknown, then the errors in $log(\dot{M} )$ 
can be as large as $\approx 1$ (in units of $\dot{M}$/yr, namely e.g. $log(\dot{M}) = -9 \pm 1$ ) 
for a COS spectrum. If the $M_{\rm wd}$ and $d$ are known, 
a high S/N COS spectrum will have errors of
barely 100 K in $T_{eff}$ (for a WD fit), or $\approx 0.2$ in $log(\dot{M})$ (e.g. $\log(\dot{M} ) = -9.0 \pm 0.2$).

Simple inversion of a Gaia parallax can provide 
an acceptable distance only when a precise parallax for
an individual object is used \citep{luri}. Uncertainties are 
typically around 0.04 mas for Gaia sources brighter than $\sim$14 mag, around 0.1 mas 
for sources with a G magnitude around 17, and around 0.7 mas at the faint end, around 
20 mag. Irregular color and brightness changes introduce 
uncertainties and must be monitored. Gaia Data Release 1  parallaxes proved successful
for virtually all of the test case CVs and the old nova RR Pic.
SS Cygni's distance of $\approx$ 114 $pc$ for the disk instability model to be valid was solidly 
confirmed with a Gaia parallax distance of 117 $pc$ \citep{ramsay}.

Schaefer \citep{Schaefer3} has concluded a study of the GAIA parallaxes of classical novae and recurrent novae
in which he assesses the reliability of the distance for each object. He ranks the GAIA results for CI Aql and 
T Pyx in his "gold" category meaning the highest level of reliability. 

\subsection{FUV Spectroscopic Modeling} \label{fuv}

\begin{figure}[h!]
\vspace{-10.cm}
\includegraphics[width=0.8\textwidth, angle =0.00]{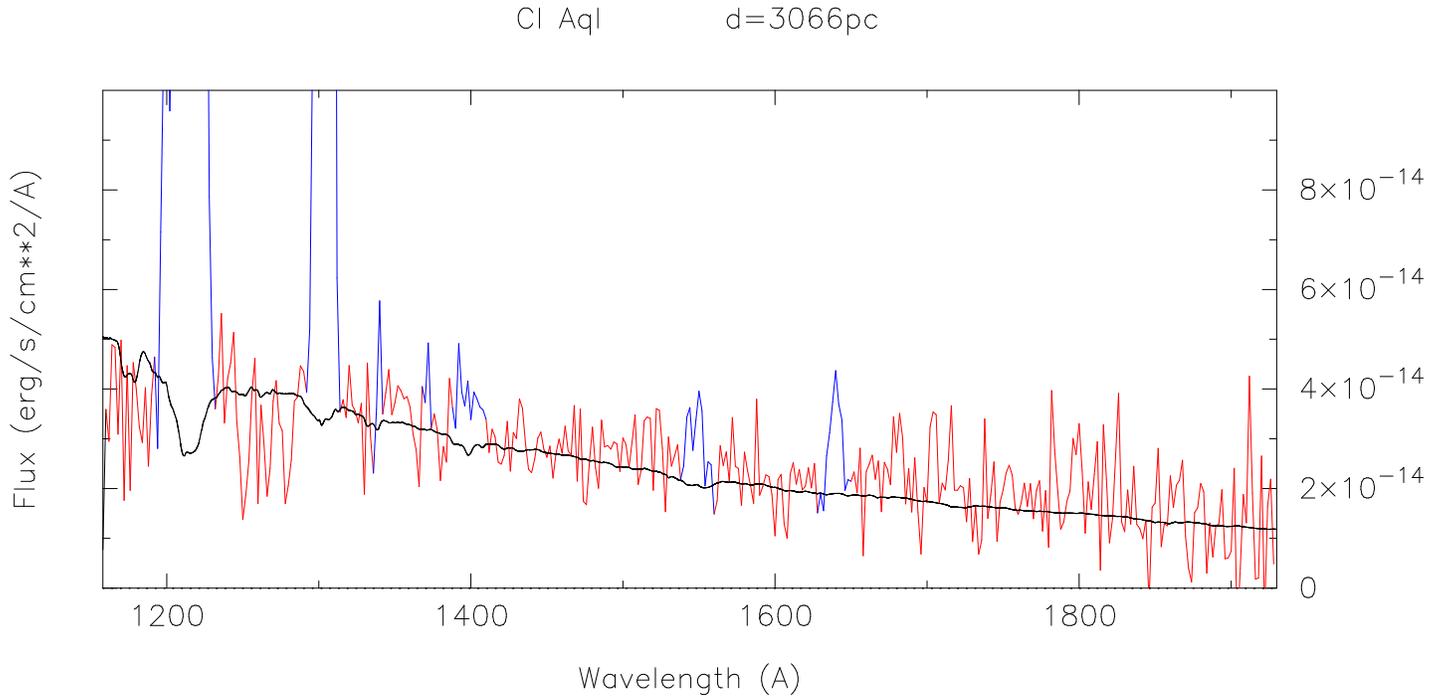}
\vspace{-1.5cm}
\caption{An accretion disk model fit to CI Aql with inner truncation. The inner disk radius was taken to be 5200 km. The outer disk radius, Rd, was chosen such that the outer disk's surface temperature reaches 
only 5,000 $K$ so its contribution to the FUV is insignificant.  
 This best fit truncated disk has an accretion rate of $dM/dt= 4\times 10^{-8} M_\sun/yr$}.
\label{ciaql_accdisk}
\end{figure}

We carried out CI Aql accretion disk fits scaled to the Gaia distance of 3.062 kpc for the 
WD mass reported in \S\ref{ephem} of 0.996 $M_\sun$, with LK's inclination of $71^{\circ}$.
$E_{B-V}$ was set to 0.8 although values as large as 0.98 
appear in the literature. We constructed disk models from scratch with fixed dimensions. 
The inner disk radius was taken to be 5200 km while the outer disk radius, Rd, was chosen such that the outer disk's surface temperature reaches 
only 5,000 K so its contribution to the FUV is insignificant. This best fit yields an accretion rate of $dM/dt= 4\times10^{-8} M_\sun/yr$, 
and its spectrum is displayed in Fig. \ref{ciaql_accdisk}.
We examined the possibility that a hot white dwarf could be contributing FUV flux, but its 
radius and surface temperature would have to be untenably large to provide the observed 
flux at a distance of 3 kpc. 

\section{Discussion \& Conclusions}

We have characterized FUV spectrum of the key RNe, CI Aql.  For CI Aql, with a Gaia distance of 3.062 kpc, an accretion disk model satisfying the new distance yields an accretion rate $4\times 10^{-8} M_\sun/yr$ for $M_{\rm wd} = 1.0 M_\sun$, with a white dwarf radius of 5200 km. 

\citet{shara} derived WD masses and accretion rates for all of the known recurrent novae from 
relationships between WD masses and accretion rates versus nova characteristics such as observed outburst 
amplitudes, decline times, and envelope masses. Their grids of multicycle nova evolution cover a wide range of accretion rates and white dwarf masses.
Their white dwarf masses and accretion rates are, for 
T Pyx, 1.23 $M_\sun$, 1.12 $\times 10^{-7} M_\sun$/yr; for IM Nor, 1.21 $M_\sun$, 4.8 $\times 10^{-8} M_\sun$/yr; 
and for CI Aql, 1.21 $M_\sun$, 1.12 $\times 10^{-7}$ $M_\sun$/yr. 
By comparison, \citet{godon18} 
derived $\dot{M} \sim 10^{-7}$ $M_\sun$/yr and $M_{\rm wd} = 1.2 M_\sun$ for T Pyx, which is in good agreement. For CI Aql we derived 4 $\times 10^{-8} M_\sun$/yr, 
which is nearly a factor of three smaller than \citet{shara}.

WH found the WD in CI Aql to show a marginally significant net mass gain, $\Delta M_1$, over its full outburst
and recovery, via simultaneous solution of light curves and eclipse timings for ephemeris quantities
and our revisit of CI Aql finds the same pre- vs. post outburst $\Delta M_1$ as in
WH, with RVs from double-lined spectra added to the database.
$\Delta M_1$'s accuracy continues to be set entirely by the less numerous pre-outburst
data. If the pre-outburst uncertainty in orbit period ($1.4\times 10^{-6}$ days) had been as 
small as its post-outburst uncertainty ($9.4\times 10^{-8}$ days), $\Delta P$'s overall standard
error would have been about 20 times smaller, as would that of $\Delta M_1$, which scales with $\Delta P$.
The next outburst should produce more definitive results if followed intensively.

For T Pyx, prototype of the short period RNe subclass, \citet{patterson} found the WD to be 
losing mass because the ejected mass exceeds the accreted mass over the time between outbursts ($\Delta t_{ob}$),
with accreted mass estimated as $dM_1/dt$ $\times$ $\Delta t_{ob}$. 
Contrasting observational and astrophysical circumstances between T Pyx and 
CI Aql suggest caution in making comparisons. For example, T Pyx's light curve data \citep{patterson}
are much more numerous than CI Aql's and better distributed in time 
in regard to pre- and post-outburst coverage. 
These points lead to a very strong $dP/dt$ measurement for T Pyx (the basis for \cite{patterson}'s
$dM_1/dt$ estimate) rather than our consistently negative but only $2.6\sigma$ $dP/dt$ result for CI Aql.
That CI Aql's $dP/dt$ could be measured at all, after previous attempts had 
failed, is due to the enhanced accuracy of the
ephemeris-measuring algorithm applied by WH and in this paper's \S\ref{ephem}. 
Note that the T Pyx masses -- especially 
that of the lobe-filling donor star -- differ greatly from those of CI Aql, and the 
system dimensions are also very different, thus complicating the comparison. 

For CI Aql's WD to have gained mass
through its recent eruption event would require large-scale and nearly impulsive accretion
when its $dP/dt$ was not measured, some time within the overall outburst process.
However that interval is when light curves are disturbed so that measurements of $P$ and corresponding white
dwarf mass estimates are unreliable.
Given these considerations, we regard the simple period differencing (pre- vs. post outburst)
of \S\ref{simple} as the preferred way to estimate $\Delta M_1$ where suitable data exist.
Notably the sign of CI Aql's $\Delta M_1$ has been positive in all of our computations. Note that the WD masses of both IM Nor and CI Aql are far larger than the canonical single white dwarf mass of 
$0.6 M_{\odot}$. 
This fact in itself suggests that the WD in CI Aql may be growing mass. Whether all, some, or none of the WDs in short period
RNe are expected to produce SN Ia's remains an active issue. With only three such systems now known, the score is tied at 1 to 1 in regard to conclusions for and against these rare objects heading toward an ultimate catastrophe.

\acknowledgments

This research was supported by HST grant GO14641 to Villanova University. SS acknowledges 
partial support to ASU from various NASA grants. PG wishes to thank William P. Blair for 
his kind hospitality at the Rowland Department of Physics and Astronomy, the Johns Hopkins University, Baltimore, MD.
We thank D. Sahman for sending the RVs and for thoughtful responses to questions.

\clearpage

\

\clearpage

\clearpage

\begin{deluxetable}{ccccccccc}
\tablewidth{0.0pt}
\tablecaption{Basic Observed Parameters of CI Aql \label{params}}
\tablehead{\colhead{$V_{max}$} & \colhead{$V_{min}$} & \colhead{$t_3$} &  \colhead{$P_{orb}$ $(days)$} & \colhead{$E_{B-V}$} & \colhead{$i$} & \colhead{$M_{\rm wd}$ $(M_{\odot})$} & 
\colhead{$d$ $(kpc)$} & \colhead{Nova Outbursts}}  
\startdata

  9.0    &   16.7  &    32 &   0.62   & 0.8 & 71 deg & 0.98 $M_\sun$ & 3.06 & 1917, 1941, 2000 
\enddata
\tablecomments{Item $t_3$ is the time in days past peak light for brightness to drop by $3^m$.}

\clearpage

\end{deluxetable}

\begin{deluxetable}{lcc}

\tablecaption{CI Aql Post-outburst Simultaneous Light-RV-Timing Solutions  \label{rezults}}
\tablehead{\colhead{Parameter} &  \colhead{RV \& Timing Input} & \colhead{RV, light curve, \& Timing Input}}
\startdata
$a$ ($R_\sun$) & $4.583\pm0.088$ & $4.583$  \\
$V_\gamma$ $(km\,s^{-1})$ & $+10.3\pm2.7$ & $+10.3$  \\
$m_2/m_1$ & $2.39\pm0.13$ & $2.39$ \\
$HJD_0$ & $2453652.75826\pm0.00026$ & $2453652.75821\pm0.00018$ \\
$P_0$ & $0.61836041\pm0.00000012$ & $0.618360142\pm 0.000000094$ \\
$dP/dt$ & $0.00$ & $-2.49\pm0.95 10^{-10}$ \\
$M_1/M_\sun$ & $0.996$ & \ldots  \\
$M_2/M_\sun$ & $2.38$ & \ldots  \\
$R_2/R_\sun$ & $2.10$ & \ldots  \\
\enddata
\tablecomments{Quantities in the last column without standard errors were adopted
from the middle column. The orbital inclination was $69^\circ$ in all solutions.
The mean donor star radius, $R_2/R_\sun$, follows from the lobe filling condition
for synchronous rotation. The middle column's $dP/dt$ is reported as zero since
deletion of light curve input left $dP/dt$ non-significant.}

\end{deluxetable}

\end{document}